\documentclass{edp-conf}
\usepackage{graphicx}
\begin{document}

\TitreGlobal{SF2A 2002}

\title{Molecular gas in the 3-ringed Seyfert/Liner galaxy NGC 7217} 
\author{Combes F.}
\address{LERMA, Observatoire de Paris, France}
\author{Garc\'ia-Burillo S.}
\address{Observatorio Astronomico Nacional, Madrid, Spain}
\author{Boone F.}
\address{LERMA, Observatoire de Paris, France}
\author{Hunt L.K.}
\address{Instituto di Radioastronomia/CNR Firenze, Italy}
\author{Leon S., Eckart A.}
\address{I. Physikalisches Institut, Universitaet zu Koeln, Germany}
\author{Baker A.J., Tacconi L.J., Englmaier P.}
\address{MPE, Garching, Germany}
\author{Schinnerer E.}
\address{Caltech, Pasadena, California, USA}
\author{Neri R.}
\address{IRAM, Grenoble, France}

\runningtitle{Molecular gas in NGC 7217}
\setcounter{page}{237}
\index{Combes, F.}
\index{Garc\'ia-Burillo, S.}
\index{Boone, F.}
\index{Hunt, L.K.}
\index{Leon, S.}
\index{Eckart, A.}
\index{Baker, A.J.}
\index{Tacconi, L.J.}
\index{Englmaier, P.P.}
\index{Schinnerer, E.}
\index{Neri, R.}

\maketitle
\begin{abstract}
We present CO(1-0) and CO(2-1)  maps of the Seyfert galaxy NGC 7217,
obtained with the IRAM interferometer, at 3" and 1.5" resolution
respectively. The nuclear ring (at r=12"=0.8kpc) is predominant in the CO maps, 
with a remarkable surface density gradient between the depleted region inside the ring
and the inner border of the ring.
The CO nuclear ring is significantly broader (500-600pc) than the
dust lane ring. The CO(2-1)/CO(1-0) ratio is around 1, typical of optically
thick gas with high density.
The overall morphology of the ring is quite circular, with no evidence
of non-circular velocities. In the CO(2-1) map, a central concentration
might be associated with the circumnuclear ring of ionised gas detected inside r=3"
and interpreted as a polar ring by Sil'chenko and Afanasiev (2000). 
Our interpretation is more in terms of a bar/spiral structure, in 
the same plane as the global galaxy but affected by non-circular motions,
which results in a characteristic S-shape of the isovels. This nuclear 
bar/spiral structure, clearly seen in a V-I HST colour image, is essentially
gaseous and might be explained with acoustic waves.
\end{abstract}
%

\vspace{-0.3cm}
\section{Introduction}

NGC 7217 is one of the first galaxies observed from the NUGA (NUclei of GAlaxies)
sample of about 20 spirals with AGN, which are being mapped at high
resolution with the IRAM interferometer to explore
 their molecular content, and to determine their possible fueling
mechanisms (primary or nuclear bars, spiral waves, warps, or m=1 perturbations).

NGC 7217 is an (R)SA(r)ab galaxy classified as a LINER/Seyfert. It is particularly
axisymmetric, and possesses 
nuclear, inner and outer rings, at 8, 31 and 77" (Buta et al. 1995).
Dominated by a massive and extended bulge, the spiral structure is flocculent,
and a possible oval distortion might be the vestige of an ancient bar,
which was responsible for the formation of the three rings.

We have observed in 2001 this galaxy with the IRAM interferometer in the 
CO(1-0) and CO(2-1) lines, with resolution of 3.2''x2.8'' and 1.6''x1.4''
 respectively.

\vspace{-0.3cm}
\section{Preliminary results}

\begin{figure}[h]
   \centering
   \includegraphics[width=9cm]{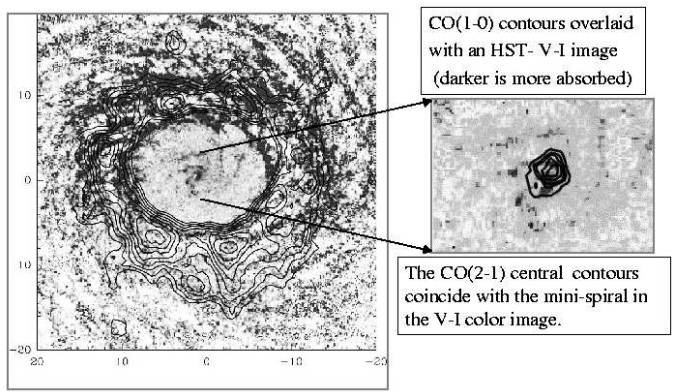}
      \caption{CO(1-0) contours superposed on the HST color image ({\it left}),
and a zoom into the center showing the CO(2-1) contours ({\it right}). }
       \label{fig1}
   \end{figure}

Fig. \ref{fig1} shows in contours the particularly regular nuclear ring
observed in the CO(1-0) line. The ring is wider in the southern part, 
and has strikingly steep contours at its inner edge. No CO(1-0) emission
is detected inside the ring, while there is a spot of CO(2-1) emission
towards the nucleus. HST colour pictures
reveal a red nuclear ring coinciding with the CO(1-0) ring.
The extinction also delineates a central spiral structure,
with which the CO(2-1) appears associated.

The CO(2-1)/CO(1-0) ratio has been computed across the map, 
by comparing only the same uv-frequencies and obtaining the two maps with
the same spatial resolution. The ratio is consistent
with 1 in most of the clumps, as for optically thick CO clouds.
The velocity field is very regular, corresponding to 
rotation, with no signature of streaming motions.


\vspace{-0.3cm}
\begin{thebibliography}{}
\bibitem{}Buta, R., van Driel, W., Braine, J., Combes, F.,
 Wakamatsu, K., Sofue, Y., Tomita, A.: 1995, ApJ 450, 593
\bibitem{}Sil'chenko O.K., Afanasiev V.L.: 2000, A\&A 364, 479
\end{thebibliography}
\end{document}